\documentclass[11pt]{article}

\usepackage{natbib}
\usepackage{amsmath,mathtools}
\usepackage[margin=1in]{geometry}
\usepackage{psfrag,epsf}
\usepackage{enumerate}
\usepackage[utf8]{inputenc}
\usepackage{amsfonts}
\usepackage{tikz}
\usepackage{xurl}
\usepackage{multirow}
\newcommand{\overbar}[1]{\mkern 1.5mu\overline{\mkern-1.5mu#1\mkern-1.5mu}\mkern 1.5mu}
\usepackage{multirow}
\usepackage{multicol}
\usepackage{bbm}
\usepackage{xr-hyper}
\makeatletter
\newcommand*{\addFileDependency}[1]{
  \typeout{(#1)}
  \@addtofilelist{#1}
  \IfFileExists{#1}{}{\typeout{No file #1.}}
}
\makeatother

\newcommand*{\myexternaldocument}[1]{%
    \externaldocument{#1}%
    \addFileDependency{#1.tex}%
    \addFileDependency{#1.aux}%
}
\myexternaldocument{Supp}
\usepackage[colorlinks=true,linkcolor=black,citecolor=black, breaklinks]{hyperref}

\usepackage{breakurl}
\usepackage{float}
\usepackage{verbatim}
\newcommand{\RNum}[1]{\uppercase\expandafter{\romannumeral #1\relax}}
\title{Bayesian sample size calculations for SMART studies}

\author{Armando Turchetta$^{1}$,
Erica E.M. Moodie $^{1}$,
David A. Stephens$^{2}$,
Sylvie D. Lambert$^{3}$\\
$^{1}$Department of Epidemiology, Biostatistics and Occupational Health \\
$^{2}$Department of Mathematics and Statistics  \\
$^{3}$Ingram School of Nursing\\
McGill University, Montreal, Québec, Canada
}
\date{}
\begin{document}
\maketitle

\label{firstpage}

\begin{abstract}
   In the management of most chronic conditions characterized by the lack of universally effective treatments, adaptive treatment strategies (ATSs) have been growing in popularity as they offer a more individualized approach, and sequential multiple assignment randomized trials (SMARTs) have gained attention as the most suitable clinical trial design to formalize the study of these strategies. While the number of SMARTs has increased in recent years, their design has remained limited to the frequentist setting, which may not fully or appropriately account for uncertainty in design parameters and hence not yield appropriate sample size recommendations. Specifically, standard frequentist formulae rely on several assumptions that can be easily misspecified. The Bayesian framework offers a straightforward path to alleviate some of these concerns. In this paper, we provide calculations in a Bayesian setting to allow more realistic and robust estimates that account for uncertainty in inputs through the `two priors' approach. Additionally, compared to the standard formulae, this methodology allows us to rely on fewer assumptions, integrate pre-trial knowledge, and switch the focus from the standardized effect size to the minimal detectable difference. The proposed methodology is evaluated in a thorough simulation study and is implemented to estimate the sample size for a full-scale SMART of an Internet-Based Adaptive Stress Management intervention based on a pilot SMART conducted on cardiovascular disease patients from two Canadian provinces.
    
\end{abstract}

\section{Introduction}
Precision medicine has  become a popular topic in the field of healthcare. Within this medical model, treatments and decision rules are personalized and tailored to  patient characteristics, shifting the focus from the traditional treatment of the diagnosis to the treatment of the patient. In settings where there is a lack of a universally effective treatment, several interventions are often needed to prevent onset and alleviate symptoms improving the patient quality of life, requiring a sequential, individualized approach whereby interventions are adapted and re-adapted over time in response to the specific needs and evolving condition of the individual.

In order to estimate the optimal individualized sequence of treatments for each patient, adaptive treatment strategies (ATSs), also known as dynamic treatment regimes (DTRs), have been introduced \citep{lavori,mur1}. To formalize the study of these regimes, the sequential multiple assignment randomized trial (SMART) has been developed \citep{lavori2,mur2}. SMARTs are based on multiple stages, each representing a clinical decision point: at each step, the patients are randomized accounting for a small set of characteristics or responses to previous interventions. When compared to standard randomized controlled trials (RCTs), SMARTs have some relevant advantages: most importantly, they allow for the direct comparison of multiple ATSs and to discover interactions between treatments. In this type of design, it is crucial  to identify carry-over effects of previous treatments on the future ones, so as to avoid the detection of interventions that only appear to be optimal in the short term but are not in fact optimal in the long term, e.g.~because they may preclude later, more effective therapies. SMARTs are designed to identify interactions that regular RCTs are likely to miss, as the latter are typically powered to make comparisons between average effects in each treatment arm. Furthermore, due to their randomization process and the variety of treatments, SMARTs are ethically advantageous and appealing to the study participants \citep{smart_ad}. So far, SMARTs have been deployed to estimate optimal strategies in a wide range of fields, such as weight loss \citep{app1}, substance abuse \citep{app4}, and cancer \citep{app2,app3}, with particular emphasis on prostate cancer \citep{app5}. Notably, because of the adaptive nature of the treatments under consideration, SMARTs have assumed an important role in the management of chronic diseases \citep{moodie_book}, namely ADHD \citep{app6}, schizophrenia \citep{app7}, and alcohol dependence \citep{app8}, among others. In this paper, for example, we will apply our proposed methodology to data from a web-based stress management program study, named Internet-Based Adaptive Stress Management Pilot SMART \citep{grant}, which employs the SMART design to overcome the necessity for interventions that are tailored to the patients' needs, which often lead to better outcomes in internet-based programs.
While the number of SMARTs has increased in recent years, and it is clear they have the potential for playing an important role in the management of chronic conditions, their theoretical features are yet to be fully discovered. Several sample size estimation methods have recently been introduced  for determining the optimal adaptive treatment strategy among all the available regimes accounting for multiple comparisons \citep{oetting, mc1, mc3, mc2}. However, since most of the primary analyses performed on SMARTs are focused on the comparison between two means or two strategies, and given that the mean outcome of a strategy is a weighted mean across outcomes of individuals whose paths are consistent with the strategy, frequentist calculations for SMARTs sample sizes with continuous outcomes are similar to traditional randomized clinical trials \citep{oetting,moodie_book2, binary}.
Despite their similarity with more classical RCTs, SMART sample size calculations generally rely on additional assumptions and specifications of key design parameters. Yet, little attention has been paid to the robustness of these calculations to model misspecification or uncertainty. In particular, response rates to initial treatments and a standardized effect size must be fixed, leading to a decrease in power if responses are misspecified or if there is variability around them. In Bayesian literature, this shortcoming is commonly known as local optimality, and the Bayesian  `two priors'  approach has been introduced as a flexible and useful extension of standard frequentist methods in order to overcome this drawback \citep{twop_1, twop_2, twop_3, twop_4, twop_5}. 

In this paper, we adapt the `two priors'  approach to the SMART design, and we analyze the performance of this Bayesian method to sizing a SMART in a detailed simulation study. With respect to the standard frequentist formulae, the proposed method relies on fewer assumptions, allows for greater flexibility in the design stage, and leads to more robust sample size calculations. The major drawback of this approach lies in the specification of the variance of the strategy means' estimator, which, contrary to frequentist calculations, needs to be specified. Although the use of crude estimates from pilot studies of the variance components needed for sample size computations is controversial \citep{sd1, sd2, sd3}, and SMART pilot studies are generally sized to ensure that a sufficient number of subjects for each treatment sequence is observed with high probability \citep{smart_n} rather than through precision-based approaches (although an alternative has recently been introduced in \cite{prec}), we propose to marginalize the Bayesian power function over the posterior distribution of the variance components estimated on pilot data in order account for their variability.

The paper is structured as follows. In Section 2, we give an overview of the frequentist sample size formula and we outline its Bayesian generalization. In Section 3, we analyze the performance of the proposed method in terms of power and type I error through an extensive simulation study. In Section 4, we apply our method to data from the Internet-Based Adaptive Stress Management Pilot SMART in order to estimate the sample size for its full-scale version. Section 5 concludes.

\section{Methodology}

\begin{figure}[H]
  \centering
\includegraphics[width=5in]{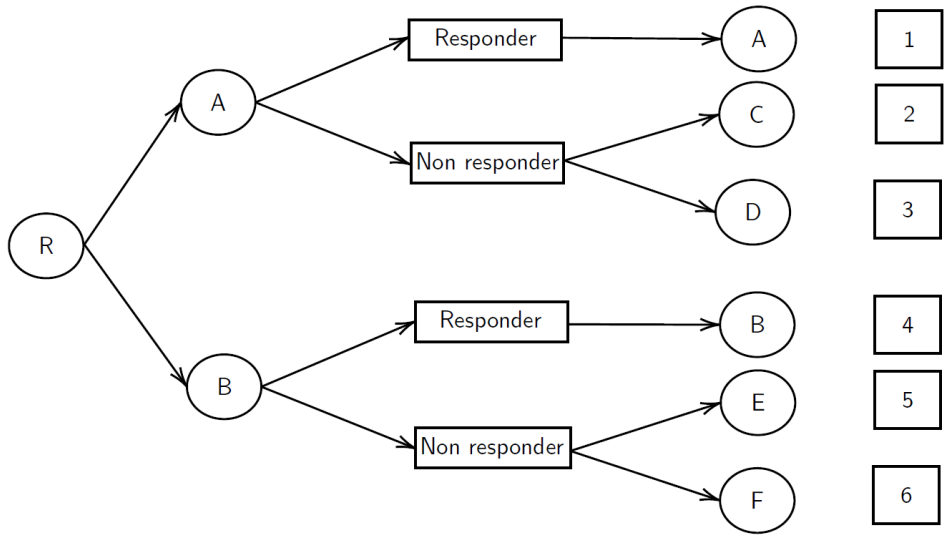}
\caption{SMART scheme }\label{fig:SMART_scheme}
\end{figure}

Let us consider a SMART with $K$ decision points; Figure \ref{fig:SMART_scheme} depicts a 2-stage example. We focus on the comparison of two adaptive treatment strategies that begin with different initial treatments. Let $S_1$ be the pre-treatment information, $A_{j}$ the treatment assigned at stage $j$, $S_j$ the intermediate outcome after treatment $A_{j-1}$ and $d_j$ the decision rule at the decision point $j$. The overbar denotes the accrual of information up to the index, e.g.~$\overbar{S}_j=\{S_1,...,S_j\}$.
The continuous outcome is denoted $Y$.

\subsection{Frequentist sample size estimate}
Using results from \cite{mur1} and  \cite{mur2}, under the assumption that at any decision point, and for any given history, the probability of any treatment included in an adaptive treatment strategy being assigned is positive, a consistent estimator of the mean outcome $Y$ under strategy $\overbar{d}_k$, which we denote $\mu_{\overbar{d}_k}$, is
\[
\widehat{\mu}_{\overbar{d}_k}= \frac{\textbf{P}_n\left[\prod_{j=1}^k \frac{I\left\{\overbar{A}_j=d_j(\overbar{S}_j,\overbar{A}_{j-1})\right\}}{\Pr(d_j|\overbar{S}_j,\overbar{A}_{j-1})}Y\right]}{\textbf{P}_n\left[\prod_{j=1}^k \frac{I\left\{\overbar{A}_j=d_j(\overbar{S}_j,\overbar{A}_{j-1})\right\}}{\Pr(d_j|\overbar{S}_j,\overbar{A}_{j-1})}\right]}
\]
where $\textbf{P}_n$ represents the sample average and $I$ the indicator function. Furthermore, defining

\[
U(\overbar{S}_k,\overbar{A}_k,{\overbar{d}_k,\mu_{\overbar{d}_k}})=\prod_{j=1}^k \frac{I\left\{\overbar{A}_j=d_j(\overbar{S}_j,\overbar{A}_{j-1})\right\}}{\Pr(d_j|\overbar{S}_j,\overbar{A}_{j-1})}(Y-\mu_{\overbar{d}_k}),
\]
a consistent estimator of the variance of $\sqrt{n}(\widehat{\mu}_{\overbar{d}_k}-\widehat{\mu}_{\overbar{d'}_k})$ is $\textbf{P}_n(U^2(\overbar{S}_k,\overbar{A}_k,{\overbar{d}_k,\mu_{\overbar{d}_k}}) + U^2(\overbar{S}_k,\overbar{A}_k,{\overbar{d'}_k,\mu_{\overbar{d'}_k}}))$ and the test statistic
\[
Z= \frac{\sqrt{n} (\widehat{\mu}_{\overbar{d}_k}-\widehat{\mu}_{\overbar{d'}_k})   }{\textbf{P}_n(U^2(\overbar{S}_k,\overbar{A}_k,{\overbar{d}_k,\mu_{\overbar{d}_k}}) + U^2(\overbar{S}_k,\overbar{A}_k,{\overbar{d'}_k,\mu_{\overbar{d'}_k}}))}
\]
is normally distributed for large samples. By writing the variance of $\sqrt{n}\widehat{\mu}_{\overbar{d}_k}$ as
\[
\tau^2_{\overbar{d}_k}=E_{\overbar{d}_k}\left[
\prod_{j=1}^k \frac{(Y-{\mu}_{\overbar{d}_k})^2}{\Pr(d_j|\overbar{S}_j,\overbar{d}_j)}
\right],
\]
if there is no pre-treatment information $S_1$ and we consider a two-stage SMART ($k=2$), indicating the intermediate outcome $S_2$ with $R$, it follows that
\[
\tau^2_{\overbar{d}_k}=E_{\overbar{d}_k}\left[\frac{(Y-\mu_{\overbar{d}_k})^2}{\Pr(a_1)\Pr(a_2|a_1,R=1)}\right]\Pr(R=1) + E_{\overbar{d}_k}\left[\frac{(Y-\mu_{\overbar{d}_k})^2}{\Pr(a_1)\Pr(a_2|a_1,R=0)}\right]\Pr(R=0).
\]

Assuming that
\begin{enumerate}
    \item the variance of the outcome $Y$, conditional  on the intermediate outcome $R$, is not greater than the variance of the strategy mean, i.e. 
    \begin{equation}\label{eq:dis}E_{\overbar{d}_k}\left[(Y-\mu_{\overbar{d}_k})^2|R\right]\leq E_{\overbar{d}_k}\left[(Y-\mu_{\overbar{d}_k})^2\right]; \end{equation}
    \item the response rates to the initial treatments are equal; and
    \item patients are randomized equally to the available treatments,
\end{enumerate}
and considering the SMART design outlined in Figure \ref{fig:SMART_scheme} where responders to the initial treatments have only one subsequent treatment option, an upper bound of $\tau^2_{\overbar{d}_k}$ is ${2\sigma^2_{\overbar{d}_k}(2-p)}$, where $p$ is the common response rate to the initial treatments and $\sigma^2_{\overbar{d}_k}$ is the marginal variance of the strategy. Considering the system of hypotheses
\begin{equation}\label{eq:hp1}
\left\{
                \begin{array}{ll}
                  \text{H}_0: \mu_{\overbar{d}_k^1}-\mu_{\overbar{d}_k^2}=0\\
                  \text{H}_1: \mu_{\overbar{d}_k^1}-\mu_{\overbar{d}_k^2}> 0
                \end{array}
              \right.
\end{equation}
where $\overbar{d}_k^1$ and $\overbar{d}_k^2$ are two strategies with a different initial treatment, and using the upper bound of $\tau^2_{\overbar{d}_k}$, for a standardized effect size $\delta=\frac{\mu_{\overbar{d}_k^1}-\mu_{\overbar{d}_k^2}}{\sigma}$ where $\sigma=\sqrt{\left(\sigma^2_{\overbar{d}_k^1}+\sigma^2_{\overbar{d}_k^2}\right)/2}$, the sample size formula is given by the value of $n$ which satisfies
\[
\Pr(Z>z_{1-\alpha}|\mu_{\overbar{d}_k^1}-\mu_{\overbar{d}_k^2}=\delta \sigma)=1-\beta,
\]
which is 
\begin{equation}\label{eq:samp}
n=\frac{(z_{\beta}+z_{\alpha})^2}{\delta^2}4[2(1-p)+p].
\end{equation}

In the next section, we will outline a Bayesian formulation of these calculations, showing how prior beliefs on the uncertainty of design parameters can be incorporated into the calculations.

\subsection{Bayesian generalization: the `two priors' approach}
In accordance with the definition of Bayesian significance given by \cite{spie},  a result is considered significant if the posterior probability that the parameter of interest $\theta$ belongs to the alternative hypothesis space $\Theta_1$ is not less than a specified threshold $1-\epsilon$, i.e.~$\text{Pr}_{\pi(\cdot|\text{Data})}(\theta\in \Theta_1)\geq 1-\epsilon$. Setting $Y_n=\widehat{\mu}_{\overbar{d}_k^1}-\widehat{\mu}_{\overbar{d}_k^2}$, $\theta={\mu}_{\overbar{d}_k^1}-{\mu}_{\overbar{d}_k^2}$, and $\tau^2=\tau^2_{\overbar{d}_k^1}+\tau^2_{\overbar{d}_k^2}$ to ease the notation, for large samples,
\[Y_n|\theta\sim \mathcal{N}\left(\theta, \frac{\tau^2}{n}\right).\]
If we consider the conjugate prior distribution for $\theta$, $\pi_0(\theta)= \mathcal{N}\left(\theta; \theta_0, \sigma^2_0\right)$, called  \textit{analysis prior}, the posterior distribution of the parameter of interest is
\[
    \pi(\theta|Y_n)= \mathcal{N}\left(\theta; \frac{\tau^2 \theta_0 + n \sigma^2_0 Y_n}{\tau^2+n\sigma^2_0},\right. \left.\left(\frac{1}{\sigma^2_0} + \frac{n}{\tau^2}\right)^{-1}     \right),
\]
and it follows that the outcome of the clinical trial is significant in the Bayesian sense if $\text{Pr}_{\pi(\cdot|Y_n)}(\theta>0)\geq 1-\epsilon$, i.e. when
\[
Y_n\geq
-\frac{z_{\epsilon} \sigma_0 \sqrt{\tau^2}
\sqrt{\tau^2+n\sigma^2_0}}{n\sigma^2_0}
-\frac{ \theta_0 \sqrt{\tau^2}}
{n\sigma^2_0}.
\]

Given that in the pre-experimental phase $Y_n$ has not been observed yet, the Bayesian power function is defined as the probability of obtaining a Bayesian significant result. To compute this probability, similarly to frequentist calculations, the standard approach consists of using the distribution of $Y_n$ conditional on a value $\theta_d$ under the alternative hypothesis.

In order to overcome the local optimality issue -- i.e. optimal performance only under specific values of the design parameters, with performance losses under alternative specifications -- the `two priors' approach entails the elicitation of a second prior distribution $\pi_d(\theta)$, called \textit{design prior}, which formalizes the uncertainty around the value of the minimal detectable difference (MDD) $\theta_d$. Setting the conjugate design prior $\pi_d(\theta)= \mathcal{N}\left(\theta; \theta_d, \sigma^2_d\right)$,
the marginal distribution of the data is
\[
m_d(Y_n)= \mathcal{N}\left(Y_n; \theta_d, \frac{\tau^2}{n}+
\sigma^2_d \right),\]
hence the Bayesian power function $\eta(n)$ can be expressed as
\begin{equation}\label{eq:bay_pow}\begin{split}
\eta(n)&=\text{Pr}_{m_d(\cdot)}\{ \text{Pr}_{\pi(\cdot|Y_n)}(\theta>0)\geq 1-\epsilon\}\\
&=\Phi\left[ \frac{1}{\sqrt{\frac{\tau^2}{n} + \sigma_d^2   } 
} 
\left(\frac{\theta_0\tau^2}{n\sigma^2_0} + \theta_d
\right.\right. + \left.\left.\frac{z_\epsilon \sqrt{\tau^2}\sqrt{\tau^2 + n\sigma^2_0}}{n\sigma_0}
\right)
\right]
\end{split}\end{equation}
and the sample size is selected as $\min\{n\in \mathbb{N} :\eta(n)>1-\beta\}$ for a given threshold $1-\beta$. Note that, if $\sigma^2_d\rightarrow 0$ and $\sigma^2_0\rightarrow \infty$, Equation \ref{eq:bay_pow} reduces to the frequentist power function which leads to the sample size Formula \ref{eq:samp} when the upper bounds of the variance of the strategy means' estimator are used.

\subsection{Accounting for variability around the variance components estimates}
A drawback of the Bayesian power function consists in the specification of $\tau^2_{\overbar{d}_k^1}$ and $\tau^2_{\overbar{d}_k^2}$. In fact, contrary to the frequentist counterpart, replacing $\tau^2_{\overbar{d}_k^1}$ and $\tau^2_{\overbar{d}_k^2}$ with their upper bounds does not lead to a simplified formula which allows us to avoid the direct specification of the variance components by specifying a standardized effect size. To overcome this pitfall and properly size a full-scale SMART, we propose the integration of prior knowledge from its pilot study. However, the direct use of a plug-in estimate of the variance components from pilot studies has been generally criticized, as it often leads to underpowered trials \citep{sd1, sd2, sd3}. In order to account for the uncertainty around the estimates of $\tau^2_{\overbar{d}_k^1}$ and $\tau^2_{\overbar{d}_k^2}$, instead of using their crude estimates, we propose the use of their posterior distribution based on pilot data to marginalize the Bayesian power function given in Equation \ref{eq:bay_pow}.

Indicating with the superscript $p$ the quantities that are pertinent to the pilot study, from the previous section we have that 
\[Y^p_n|\theta,\tau^2\sim \mathcal{N}\left(\theta, \frac{\tau^2}{n}\right).\]
A possible choice of prior conjugate distribution for $(\theta,\tau^2)$ is the Normal-inverse-chi-squared (NIX) density with parameters $\theta_p,\kappa_p, \sigma^2_p$ and $\nu_p$, where $\theta_p$ and $\sigma^2_p$ represent the prior values of $\theta$ and $\tau^2$, while $\kappa_p$ and $\nu_p$ set the strength of the prior specifications \citep{gauss}. 
This density has the form of the product between a Normal distribution and the PDF of a Noncentral chi-squared random variable, i.e.
\[\pi(\theta,\tau^2)=\mathcal{N}(\theta|\theta_p,\sigma^2_p)\chi^{-2}(\tau^2|\nu_p,\sigma^2_p).\]
It follows that the marginal posterior distribution $\pi(\tau^2|Y_n)$ is a Noncentral chi-squared density with parameters
\[
\nu_n=\nu_p + n, \qquad
\sigma^2_n=\frac{1}{\nu_n}\left[\sigma^2_p\nu_p + n\widehat{\tau^p}^2 +
\frac{n\kappa_p}{\kappa_p + n}\left(\theta_p-\widehat{\theta}^p\right)^2\right],
\]
where $\widehat{\theta}^p=\widehat{\mu}^p_{\overbar{d}_k^1}-\widehat{\mu}^p_{\overbar{d}_k^2}$ and $\widehat{\tau^p}^2=\widehat{\tau^p}^2_{\overbar{d}_k^1}+\widehat{\tau^p}^2_{\overbar{d}_k^2}$ are estimated form the pilot study.
Finally, indicating with $\eta(n;\tau^2)$ the Bayesian power function defined in \ref{eq:bay_pow}, the marginal power function is 
\[\eta^{m}(n)=\int_0^\infty \eta(n;\tau^2)\pi(\tau^2|Y^p_n)d\tau^2.\]

We will now assess the properties and performance of this methodology in a simulation study.

\section{Simulation study}\label{sim_sec}

In this section, we analyze through simulated data the sensitivity in terms of power and type I error of the proposed methodology and the existing frequentist sample size formulae for SMARTs to the misspecification of response rates, over-estimation of the standardized mean difference or minimal detectable difference, and a breach of Assumption \ref{eq:dis}. 

\subsection{Setting}

In this simulation study, we consider both continuous and binary outcomes, assuming the appropriateness of the Normal approximation in the latter case.  
Let $R$ be the intermediate response indicator to the initial treatment (0 for non-responders and 1 for responders), $A_k$ the treatment indicator at stage $k$, $Y$ the final outcome, and $\delta$ the standardized mean difference. Let  $p_{a_1}$ be the probability of response to the initial treatment $a_1$. If the final outcome is binary, $p_{a_1,a_2}$ is the probability of response to the second treatment $a_2$ when the first treatment $a_1$ fails, while subjects who have a positive reaction to the first stage intervention are considered as respondents also at the second stage. If $Y$ is continuous, the final outcome is sampled from a Normal distribution with mean $\text{E}[Y|A_1, R, A_2]$ and variance $\text{Var}[Y|A_1, R, A_2]$. Let us consider the SMART design illustrated in Figure \ref{fig:SMART_scheme}. Following the same structure of \cite{tech}, we set $\text{Var}[Y|A_1=a_1, R=r, A_2=a_2]=\zeta^2_{a_1,r,a_2}$, and we express the conditional mean as
\begin{multline*}
\text{E}[Y|A_1, R, A_2]=\phi_1 + \phi_2\mathbbm{1}_{\{A_1=a\}} + \phi_3(1-R)+\phi_4\mathbbm{1}_{\{A_1=a\}}(1-R)\\
+\phi_5\mathbbm{1}_{\{A_2=c\, \cup \,A_2=e \}}(1-R)+\phi_6\mathbbm{1}_{\{A_1=a\, \cup \,A_2=c \}}(1-R).
\end{multline*}
It follows that the sets of parameters that need to be specified in the continuous outcome setting are $\{\phi_l , l=1,\dots, 6\}$ and the group of standard deviations for the final responses $\{\zeta_{a,1,a}, \zeta_{a,0,c}, \zeta_{a,0,d}, \zeta_{b,1,b}, \zeta_{b,0,e}, \zeta_{e,0,f}\}$.
To assess the robustness of the sample size estimates to the issue of local optimality, the response rates to the first stage intervention are sampled from a Normal distribution truncated between 0 and 1 with mean $p_{a_1}$  and standard deviation values from $0$ to $\sigma_m$. If the outcome is binary, this variability is also added to the probabilities of success of the second treatment by sampling them from a truncated  Normal distribution with mean $p_{a_1,a_2}$. Moreover, we considered overestimations of the standardized mean difference $\delta$ or minimal detectable difference by values up to 25\%. The properties of the proposed Bayesian methodology are assessed for different choices of $\theta_0$, $\sigma^2_0$ and $\sigma^2_d$ and four combinations of the aforementioned sources of model misspecification. Specifically, in Setting 1 there is no misspecification, in Setting 2 the response rates present a standard deviation equal to $\sigma_m$, in Setting 3 the minimal detectable difference is overestimated by 25\%, and Setting 4 includes both types of misspecification. Throughout this simulation study, we set the type I error $\alpha$ to $0.05$ and the type II error $\beta$ to $0.1$.

Following the SMART design outlined in Figure \ref{fig:SMART_scheme}, we compare the strategies `administer A and, if there is no response, switch to C' and `administer B and, if there is no response, switch to E'. Note that these two treatment strategies are but two of the four possible strategies that are embedded within the trial. Varying the parameters of the data generating mechanism and the type of outcome, we simulate four scenarios:
\begin{itemize}
    \item Scenario 1: the final outcome is continuous and the sets of parameters used in the data generating algorithm are
    \[\begin{split}
    p_a=p_b&=0.5,\\
    \{\phi_1,\phi_2,\phi_3,\phi_4,\phi_5,\phi_6\}&=\{10, 5, -15, -3, 10,-3\},\\
    \{\zeta_{a,1,a},  \zeta_{a,0,c}, \zeta_{b,1,b},  \zeta_{b,0,e}\}&=\{2,2,2,3\},
    \end{split}\]
    and $\sigma_m$ is set to 0.05. The resulting standardized mean difference is $0.41$ and the average difference between strategy means is 2.
    \item Scenario 2: continuous outcome. The simulation parameters are 
    \[\begin{split}
    p_a=p_b&=0.7,\\
    \{\phi_1,\phi_2,\phi_3,\phi_4,\phi_5,\phi_6\}&=\{22, 5, -15, -7, 8,-3\},\\
    \{\zeta_{a,1,a},  \zeta_{a,0,c}, \zeta_{b,1,b},  \zeta_{b,0,e}\}&=\{6,6,2,3\},
    \end{split}\]
    and $\sigma_m = 0.04$, resulting in a standardized effect size of $0.27$ and a treatment effect of 2. Note that, in this scenario, Assumption \ref{eq:dis} upon which the standard frequentist sample size formula relies does not hold. 
    \item Scenario 3: binary outcome. The probabilities of response to the various treatments are \[(p_a, p_{ac}, p_b, p_{be})=(0.3, 0.4, 0.3, 0.2),\] and $\sigma_m$ is set to 0.04. The resulting standardized effect size and treatment effect are  0.28 and 0.14 respectively. 
    \item Scenario 4: binary outcome. The probabilities of response are \[(p_a, p_{ac}, p_b, p_{be})=(0.5, 0.65, 0.5, 0.5),\] leading to the standardized effect size $\delta=0.18$ and treatment effect $0.075$, while $\sigma_m$ is set to 0.02. As in the second scenario, Assumption \ref{eq:dis} does not hold. 
\end{itemize}

Using the calculations developed in \cite{smart_n}, we determined the sample size of the simulated pilot studies in order to guarantee that at least 6 individuals are observed in each treatment sequence with a $90\%$ probability, which resulted in a sample size of 66 in Scenarios 1 and 4, 114 in Scenario 2, and 64 in Scenario 3.
Finally, the type I error was assessed by sizing the full-scale trial in order to identify a minimal detectable difference between strategy means of 2 points in the continuous outcome setting and 0.14 points in the binary outcome setting when there is indeed no difference. 
The results are based on 3000 data replications.

\subsection{Results}

Partial results of the simulation study are presented below. Specifically, Table \ref{power_res3_1} shows the simulated power of the existing frequentist sample size formula  and Table \ref{alpha_cont}  displays the simulated type I error generated under the proposed methodology, both in the continuous outcome setting.
The full results of the simulation study are presented in the Web Appendix. Web Tables S1-S2 and S3-S4 show the results related to the proposed Bayesian formula in Scenarios 1 and 2 respectively, whereas Web Tables S6-S7 and S8-S9 provide the results for Scenarios 3 and 4. Web Table S5 depicts the performance of the frequentist formula in terms of power in the binary outcome setting. Finally, Figure \ref{fig:heat} presents a comparison in terms of power between the frequentist and Bayesian formula and Web Table S10 shows the simulated type I error in the binary outcome setting.

\subsubsection{Power}
As we can see from Table  \ref{power_res3_1}, the frequentist sample size formula performed well under the best-case scenario where there is no model misspecification and Assumption \ref{eq:dis} is not violated (Scenario 1, top-left corner), nearing the desired 0.9 power level. However, its performance quickly deteriorates when the degree of model misspecification increases, causing power to fall to 0.82 when the standard deviation of the response rates reaches 0.05, 0.77 when the $\delta$ is overestimated by 25\%, and 0.72 when both sources of model misspecification are present. 

\begin{table}[!htbp] \centering 
  \caption{Simulated power under the frequentist calculations in Scenarios 1 (left) and 2 (right) for different degrees of model misspecification.} 
  \label{power_res3_1} 
\begin{tabular}{@{\extracolsep{-0pt}} ccccccccccccccc} 
\\[-1.8ex]\hline \hline
&&&\multicolumn{6}{c}{Scenario 1} && \multicolumn{5}{c}{Scenario 2} \\\cline{4-9}\cline{11-15}\\[-1.8ex]
&&&\multicolumn{12}{c}{Response SD} \\
&& & 0 & 0.01 & 0.02 & 0.03 & 0.04 &0.05 &&0 & 0.01 & 0.02 & 0.03 & 0.04 \\ 
\hline 
\multirow{5}{3em}{\% bias of $\delta$}&0 && $0.89$ & $0.89$ & $0.88$ & $0.88$ & $0.84$ & $0.82$ & & $0.83$ & $0.83$ & $0.81$ & $0.78$ & $0.76$ \\ 
&5 && $0.88$ & $0.87$ & $0.86$ & $0.84$ & $0.82$ & $0.81$ & & $0.81$ & $0.80$ & $0.78$ & $0.77$ & $0.74$\\ 
&10 && $0.84$ & $0.84$ & $0.83$ & $0.82$ & $0.82$ & $0.79$ && $0.78$ & $0.78$ & $0.76$ & $0.75$ & $0.71$\\ 
&15 && $0.82$ & $0.83$ & $0.81$ & $0.80$ & $0.79$ & $0.76$  && $0.74$ & $0.75$ & $0.73$ & $0.72$ & $0.70$\\ 
&20 && $0.78$ & $0.78$ & $0.79$ & $0.77$ & $0.76$ & $0.74$ && $0.71$ & $0.71$ & $0.71$ & $0.69$ & $0.69$\\ 
&25 && $0.77$ & $0.76$ & $0.76$ & $0.75$ & $0.72$ & $0.72$ && $0.68$ & $0.69$ & $0.66$ & $0.65$ & $0.64$  \\ 
\hline \\[-1.8ex] 
\end{tabular} 
\end{table} 

The sample size estimates in this setting ranged from $302$ under no model misspecification to $194$ when the standardized mean difference is overestimated by $25\%$. In the second scenario we notice a similar trend, however, the decrease in power is more evident because of the violation of Assumption \ref{eq:dis}, which causes power to fall to 0.83 even in the absence of misspecification. In this scenario, the sample size estimates spanned from $628$ under no model misspecification to $402$ for the maximum misspecification of $\delta$.\\

On the other hand, the proposed Bayesian methodology provides us with the tools to offset the decrease in power caused by variability around response rates or overestimation of the standardized effect size/treatment effect. Web Tables S1 and S2 provide the results of the simulation study under Scenario 1 when the the mean of the analysis prior $\theta_0$ is set to 0 and $\widehat{\theta}^p$ respectively. It is easily noticeable that the power level is independent of the choice of the analysis prior parameters $\theta_0$ and $\sigma_0$, which, as expected, only affect the sample size. Specifically, as $\sigma_0$ decreases, $n$ increases under the neutral analysis prior centered at $0$, and decreases if $\pi_0$ is centered at the treatment effect estimated via pilot data. When no variability around the minimal detectable difference is considered, i.e. $\sigma_d=0$, the Bayesian formula generally leads to the same level of power of its frequentist  counterpart in the four settings considered (which correspond to the `corners' of Table \ref{power_res3_1}), and similar average sample size values. However, the simulation results show how the addition of variability around the minimal detectable difference via the design prior $\pi_d$ effectively mitigates the loss of power which affects the frequentist formula when the model is misspecified, generating sample size estimates that are more robust.

Similarly, Web Tables S3 and S4 provide the results of the simulation study under the second scenario. As in Scenario 1, the increase of $\sigma_d$ generates estimates that are more robust to the overestimation of the treatment effect or variability around response rates. Additionally, since the Bayesian formula does not depend on Assumption \ref{eq:dis}, even if no variability around the minimal detectable difference is considered ($\sigma_d=0$), it leads to a higher level of power with respect to the frequentist methodology, nearing the 0.9 level under no misspecification. Furthermore, in this specific setting, under a non-informative analysis prior the estimated average sample size is 741, which is $18\%$ higher than the frequentist estimate, suggesting that the frequentist formula can potentially lead to underpowered studies when Assumption \ref{eq:dis} is violated.\\

\begin{figure}[H]
  \centering
\includegraphics[width=\linewidth]{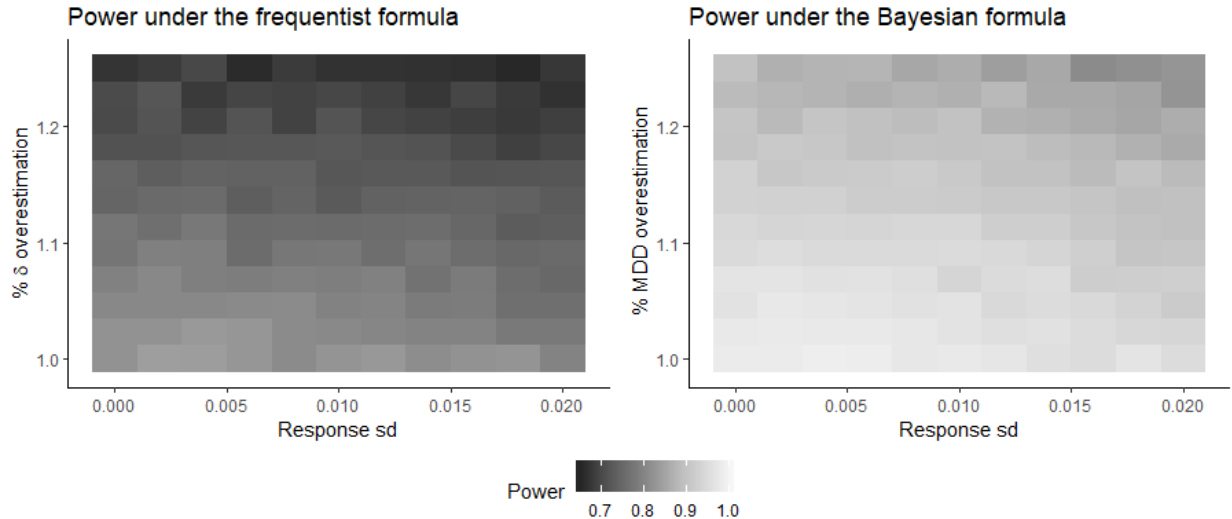}
\caption{Heatmaps representing the simulated power under the frequentist and Bayesian methodology in Scenario 4 for several degrees of model misspecification. For the Bayesian formula, the prior mean $\theta_0$ is set to $0$, $\sigma_0=100$ and $\sigma_d=0.03$. }\label{fig:heat}
\end{figure}

Analogous considerations can be made in the settings which entail a binary final outcome, as the performances of both the frequentist and Bayesian methods under Scenarios 3 and 4 respectively mirror the ones under Scenarios 1 and 2. Web Table S5 displays the performance of the frequentist formula in Scenarios 3 and 4, whereas the results related to the Bayesian methodology are presented in Web Tables S6-S7 and S8-S9.
In the frequentist setting, incrementing the degree of misspecification of the standardized mean difference, sample size estimates ranged from 728  to 466 in the third scenario and from 1516 to 912 in the fourth scenario.
A partial representation of comparison of power between the two methods is depicted in the heatmaps of Figure \ref{fig:heat}, where the Bayesian methodology is assessed for a non-informative analysis prior and $\sigma_d$ is set to 0.03.

Finally, it is important to notice that the variability of the sample size estimates generated under the proposed methodology is higher in the scenarios where Assumption \ref{eq:dis} does not hold. In fact, for example, considering the empirical distribution of the sample size estimates under each combination of prior parameters, under Scenario 4 the third quartile is on average 43\% higher than the first quartile, whereas in Scenario 1 this difference amounts to 21\%.

\subsubsection{Type I error}

Table \ref{alpha_cont} show the sensitivity of type I error of the Bayesian formula under several prior specifications in the continuous outcome settings. As expected, the simulated type I error generally attains the desired 0.05  level when $\theta_0$ is set to $\widehat{\theta}^p$, and it decreases below that threshold as $\sigma_0$ decreases when the analysis prior $\pi_0$ is centered at 0.

\begin{table}[!htbp] \centering 
  \caption{Simulated type I error under the proposed methodology for different choices of prior parameters in the continuous final outcome setting.} 
  \label{alpha_cont} 
\begin{tabular}{@{\extracolsep{6pt}} cccccccc} 
\\[-1.8ex]\hline 
\hline 
&&&&\multicolumn{4}{c}{$\sigma_d$} \\
&&& & 0 & 0.2 & 0.5 & 0.8  \\ 
\hline \\[-1.8ex] 
\multirow{4}{3.2em}{$\theta_0=0$}&\multirow{4}{1em}{$\sigma_0$}&100&& $0.042$ & $0.038$ & $0.043$ & $0.040$ \\ 
&&3 && $0.036$ & $0.035$ & $0.039$ & $0.040$ \\
&&2& & $0.037$ & $0.035$ & $0.040$ & $0.037$ \\
&&1 && $0.017$ & $0.022$ & $0.021$ & $0.030$ \\ 
\\[-1.8ex] 
\hline\\[-1.8ex] 
\multirow{4}{3.2em}{$\theta_0=\widehat{\theta}^p$}&\multirow{4}{1em}{$\sigma_0$}&100 && $0.046$ & $0.040$ & $0.043$ & $0.041$ \\ 
&&5 && $0.042$ & $0.040$ & $0.038$ & $0.043$ \\
&&4 && $0.045$ & $0.034$ & $0.040$ & $0.038$ \\  
&&3 && $0.040$ & $0.040$ & $0.035$ & $0.040$ \\ 
\hline \\[-1.8ex] 
\end{tabular} 

\end{table} 

Web Table S10 provides the same information in the binary outcome setting, and the results lead to the same conclusions.

\section{Application to the Internet-Based Adaptive Stress Management Pilot SMART}
The Internet-Based Adaptive Stress Management Pilot  is a pilot SMART whose goal is to inform the planning of the subsequent larger full-scale study \citep{grant}. The objective of this clinical trial is the evaluation of adaptive internet-based stress management interventions among adults with a cardiovascular disease and mild to severe levels of stress as measured by the Depression, Anxiety, and Stress Scale (DASS) \citep{dass}.
The DASS  is a set of self-reported scales aimed at the assessment of the level of depression, anxiety, and stress. Fourteen items are dedicated to each of the three conditions, resulting in three separate scores that range from 0 to 42. The stress scale evaluates difficulty to relax, nervous arousal, irritability, impatience, and agitation. Subjects with a score of 16 or higher were deemed eligible for this trial and, after 6 weeks of the first stage intervention, participants whose score fell below this threshold or improved by at least $50\%$ with respect to their baseline assessment were considered as responders.

In order to guarantee a 90\% probability of observing at least 4 patients in each sequence of treatments, 59 patients were enrolled and randomized to either a self-directed web-based stress management program or the same intervention with the addition of the assistance of a lay coach. In accordance with the SMART scheme outlined in Figure \ref{fig:SMART_scheme}, after 6 weeks responders to the first stage intervention continued with the same program, whereas non-responders were randomized to their second stage interventions, which for both arms consisted of the continuation of the first treatment or the switch to a motivational interviewing based program. For the illustrative purposes of this section, we consider the two adaptive treatment strategies $\overbar{d}_k^1$ and $\overbar{d}_k^2$ where
\begin{itemize}
    \item $\overbar{d}_k^1=$ assign the `website only' intervention and, if the patient does not respond, switch to the motivational interviewing based program;
    \item $\overbar{d}_k^2=$ assign the `website + coach' intervention and, if the patient does not respond, switch to the motivational interviewing-based program.
\end{itemize}

\begin{figure}[H]
  \centering
\includegraphics[scale=0.6]{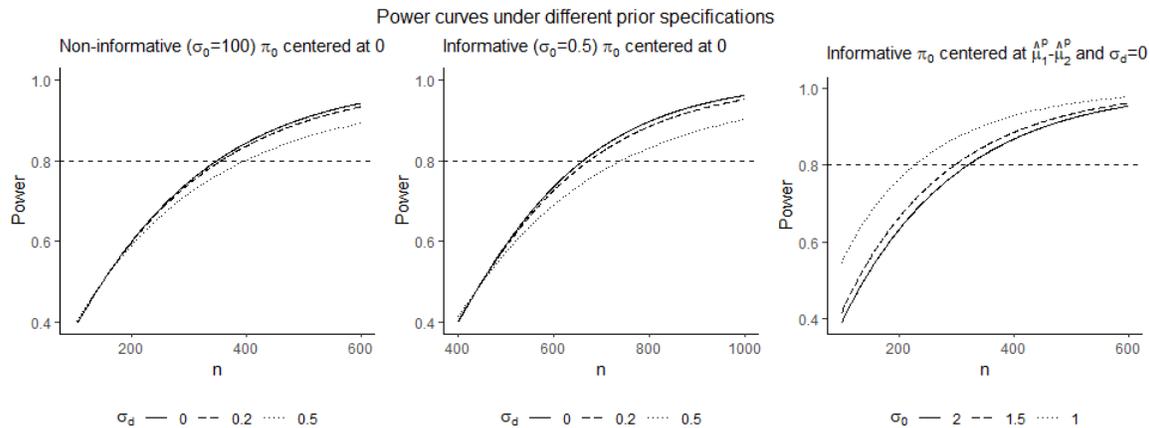}
\caption{Power curves of the full-scale version of the Internet-Based Adaptive Stress Management SMART under different choices of prior parameters:  non-informative analysis prior and varying standard deviation of the design prior (left), neutral informative analysis prior and varying standard deviation of the design prior (center), informative analysis prior centered at $\widehat{\theta}^p$ under different standard deviation values (right).}\label{fig:case_study}
\end{figure}

The full-scale version of the trial is sized to allow for the detection of a difference of 2 points in the DASS stress scale in favour of $\overbar{d}_k^1$ with $80\%$ power. The system of hypotheses is the following:
\begin{equation}\label{eq:hp2}
\left\{
                \begin{array}{ll}
                  \text{H}_0: \mu_{\overbar{d}_k^1}-\mu_{\overbar{d}_k^2}=0\\
                  \text{H}_1: \mu_{\overbar{d}_k^1}-\mu_{\overbar{d}_k^2}< 0.
                \end{array}
              \right.
\end{equation}
Figure \ref{fig:case_study} shows the power curves relative to the sample size calculations of the Internet-Based Adaptive Stress Management SMART under different specifications of the prior parameters of $\pi_0$ and $\pi_d$ and assuming the set of hyperparameters $(\theta_p,\kappa_p, \sigma^2_p, \nu_p)=(0, 1,0.1,5)$.

Using a non-informative prior $\pi_0$, the estimated sample size to identify a difference of 2 points in the DASS stress scale between strategies $\overbar{d}_k^1$ and $\overbar{d}_k^2$ with $80\%$ power is 349. If uncertainty around the minimal detectable difference is added through the standard deviation $\sigma_d$ of the design prior $\pi_d$, the required sample size increases to 357 and 399 for $\sigma_d=0.2$ and $\sigma_d=0.5$ respectively. On the other hand, if we are willing to borrow further information from pilot data, centering $\pi_0$ at the treatment effect estimated in the pilot study reduces the sample size to 321 if $\sigma_0=2$, 298 if $\sigma_0=1.5$, and 228 if $\sigma_0=1$.

Since the frequentist methodology requires additional assumptions on the response rate to the initial treatments and is based on the specification of the standardized mean difference rather than the treatment effect, a natural counterpart to the Bayesian sample size estimations is not achievable in real data applications. Assuming that Assumption \ref{eq:dis} is not violated and setting a $40\%$ probability of response to the first stage interventions, the sample size estimates under the frequentist methodology for a standardized mean difference of $0.20$, $0.30$, and $0.5$ are $990$, $440$, and $160$ respectively.

\section{Discussion}

In this paper, we outlined a Bayesian extension to frequentist sample size formulae for SMARTs which relies on fewer assumptions and ensures more flexibility in the specification of key design parameters. The application of the `two priors' approach to the framework of SMARTs allows us to (1) account for variability around the minimal detectable difference generating more reliable estimates and (2) integrate pre-trial knowledge via the analysis prior $\pi_0$. Through a simulation study we demonstrated that, with respect to its frequentist counterpart, this methodology generally leads to sample size estimates that are more robust to model misspecification in terms of power. Additionally, borrowing pre-trial knowledge from pilot data through the elicitation of the analysis prior $\pi_0$ is a useful tool to decrease the sample size without compromising the power of the full-scale study.
Furthermore, the marginalization of the Bayesian power function over the posterior distribution of the variance components estimated from pilot data ensures that the proposed methodology does not depend on the frequentist assumptions regarding the conditional variance of the outcome and the specification of intermediate response rates to initial treatments, which can be easily misspecified. Although pilot SMARTs are generally not sized to ensure precise estimates of the variance components, since the variability around them is encapsulated in their posterior distribution, this methodology is not compromised by the risk of underpowered full-scale trials that arises from the crude estimation of variance components from pilot data. However, this procedure makes the sample size estimates subject to variability, and the simulation study showed that in certain scenarios the level of variability across data replications can be considerable.
It should be noted the proposed methodology generated sample size estimates that were on average higher than the frequentist estimations under a non-informative (or neutral) analysis prior. However, this increment is generally due to the greater assurance of the full-scale trial reaching the desired level of power that this methodology offers and, as we showed in the simulation study, it can be a consequence of the violation of the frequentist assumption on the conditional variance of the outcome, which can lead to underpowered full-scale trials under the frequentist estimates.
Moreover, some limitations in connection with the choice of prior parameters need to be highlighted. The proposed Bayesian methodology entails a certain level of subjectivity in the choice of hyperparameters. Although the shift of focus from the standardized effect size to the absolute magnitude of the treatment effect might give a more straightforward course of action to elicit the prior distributions, we showed through the simulation study and the sizing of the full-scale version of the Internet-Based Adaptive Stress Management Pilot SMART  that sample size estimates and their properties vary substantially across different choices of hyperparameters. Therefore, a significant level of consideration and, eventually, a sensitivity analysis aimed at the selection of prior parameters are advised. 
Finally, in this paper, we focused on the simple SMART design with a continuous outcome where responders to the initial treatment are not re-randomized. A generalization of this methodology to other designs would require adjustments to the estimator of the strategy mean and its variance, but the Bayesian framework would remain generally similar. Furthermore, although we showed how the Normal approximation leads to satisfactory results when the final outcome is binary, the ad hoc extension of this methodology to handle binary outcomes is an interesting avenue for further developments.

\section*{Data availability statement}
The data that support the findings of this study are not shared as restrictions apply to the availability of these data, which were used under license for this study.

\section*{Supporting Information}
The Web Appendix and Tables referenced in Section \ref{sim_sec}  have been made available with this manuscript.

\bibliographystyle{apalike}
\bibliography{bibliography}

\label{lastpage}

\end{document}


\label{firstpage}

\maketitle

\section*{Web Appendix: simulation study results}
In this supplementary section we showcase the full results of the simulation study. Tables \ref{cont_bss_1_ni}-\ref{cont_bss_1_i} show the properties of the proposed Bayesian sample size methodology in Scenario 1 when the analysis prior mean $\theta_0$ is set to $0$ and to the treatment effect estimated via pilot data $\widehat{\theta}^p$ respectively. Tables \ref{cont_bss_2_ni}-\ref{cont_bss_2_i},  \ref{bss_1_ni}-\ref{bss_1_i}, and \ref{bss_2_ni}-\ref{bss_2_i} provide the same information for Scenarios 2,  3, and 4. Table \ref{power_res3_2} depicts the performance of the frequentist formula in terms of power in the binary outcome setting.
Finally, Table \ref{alpha_bin} illustrates the simulated type I error in the binary outcome setting under the Bayesian methodology.

\begin{sidewaystable} \centering 
\caption{Power and average sample size (first and third quartile in brackets) under Scenario 1 computed using the Bayesian `two priors' approach for different values of $\sigma_0$ and $\sigma_d$. The analysis prior mean $\theta_0$ is set to $0$.}
  \label{cont_bss_1_ni} 
  \small
\begin{tabular}{@{\extracolsep{-2pt}}ccccccccccc} 
\\[-1.8ex]\hline 
\hline \\[-2.7ex] 
&&&\multicolumn{8}{c}{$\sigma_d$}\\
&&\multicolumn{2}{c}{0}&\multicolumn{2}{c}{0.2}&\multicolumn{2}{c}{0.5}&\multicolumn{2}{c}{0.8}\\\cline{3-11}\\[-1.8ex]
Setting&$\sigma_0$& Power & $n$ & Power & $n$ &Power & $n$&Power & $n$ & \\
\hline\\[-1.8ex] 
\multirow{4}{*}{1}&    100& 0.88 & 299 (263, 333) & 0.89 & 312 (276, 347) & 0.94 & 382 (335, 429) & 0.99 & 593 (519, 663) \\ 
&3&0.89 & 311 (271, 346) & 0.90 & 322 (283, 360) & 0.94 & 395 (347, 439) & 0.99 & 605 (528, 675)  \\
&2&0.90 & 321 (283, 357) & 0.90 & 333 (291, 373) & 0.94 & 407 (357, 455) & 0.99 & 617 (544, 687) \\
&1&0.88 & 366 (321, 410) & 0.89 & 379 (332, 424) & 0.94 & 462 (407, 514) & 0.99 & 682 (599, 760) \\
\\[-1.8ex]  \hline\\[-1.8ex] 
\multirow{4}{*}{2}&  100&0.83 &  301 (262, 336) & 0.84 &  310 (270, 347) & 0.88 &  381 (333, 425) & 0.94 &  590 (516, 660) \\ 
&3&0.83 &  307 (269, 344) & 0.82 &  319 (278, 355) & 0.87 &  390 (342, 436) & 0.94 &  598 (524, 667) \\
&2&0.82 &  318 (276, 357) & 0.83 &  329 (287, 367) & 0.87 &  401 (348, 450) & 0.94 &  611 (536, 688) \\
&1&0.81 &  366 (322, 406) & 0.81 &  379 (329, 425) & 0.88 &  457 (397, 513) & 0.94 &  676 (594, 753) \\
\\[-1.8ex]  \hline\\[-1.8ex] 
\multirow{4}{*}{3}&  100&0.73  & 193 (169, 214) & 0.76  & 198 (174, 220) & 0.80  & 225 (196, 250) & 0.87  & 289 (252, 322)  \\  
&3&0.74  & 201 (176, 224) & 0.77  & 206 (181, 231) & 0.81  & 233 (203, 260) & 0.89  & 302 (265, 335)  \\ 
&2&0.73  & 212 (185, 236) & 0.75  & 217 (191, 243) & 0.80  & 244 (214, 273) & 0.87  & 313 (273, 350) \\
&1&  0.72  & 254 (222, 283) & 0.74  & 262 (228, 291) & 0.81  & 293 (258, 326) & 0.88  & 366 (319, 411) \\   
\\[-1.8ex]  \hline\\[-1.8ex]  
\multirow{4}{*}{4}&  100&0.72 &  192 (167, 213) & 0.70 &  196 (171, 220) & 0.74 &  224 (196, 249) & 0.82 &  288 (252, 321) \\ 
&3&0.70 &  200 (175, 222) & 0.70 &  206 (180, 229) & 0.75 &  231 (201, 259) & 0.81 &  298 (262, 332) \\
&2&0.69 &  211 (184, 235) & 0.70 &  214 (187, 241) & 0.76 &  245 (213, 274) & 0.83 &  313 (273, 350) \\ 
&1&0.68 &  253 (223, 283) & 0.67 &  258 (224, 288) & 0.73 &  291 (254, 325) & 0.81 &  366 (318, 411) \\
\\[-1.8ex] \hline \\[-1.8ex] 
\end{tabular} 
\end{sidewaystable}

\begin{sidewaystable} \centering 
\caption{Power and average sample size (first and third quartile in brackets) under Scenario 1 computed using the Bayesian `two priors' approach for different values of $\sigma_0$ and $\sigma_d$. The analysis prior mean $\theta_0$ is set to $\widehat{\mu}^p_{\overbar{d}_k^1}-\widehat{\mu}^p_{\overbar{d}_k^2}$ (estimated from the simulated pilot study).}
  \label{cont_bss_1_i} 
  \small
\begin{tabular}{@{\extracolsep{-2pt}}ccccccccccc} 
\\[-1.8ex]\hline 
\hline \\[-2.7ex] 
&&&\multicolumn{8}{c}{$\sigma_d$}\\
&&\multicolumn{2}{c}{0}&\multicolumn{2}{c}{0.2}&\multicolumn{2}{c}{0.5}&\multicolumn{2}{c}{0.8}\\\cline{3-11}\\[-1.8ex]
Setting&$\sigma_0$& Power & $n$ & Power & $n$ &Power & $n$&Power & $n$ & \\
\hline\\[-1.8ex] 
\multirow{4}{*}{1}&    100& 0.88 & 301 (264, 336) & 0.89 & 311 (273, 345) & 0.94 & 383 (337, 425) & 0.99 & 592 (517, 662) \\
&5&0.89 & 293 (257, 328) & 0.90 & 304 (264, 340) & 0.94 & 374 (329, 417) & 0.99 & 578 (507, 646)   \\
&4&0.90 & 288 (251, 324) & 0.90 & 298 (258, 334) & 0.94 & 367 (321, 413) & 0.99 & 564 (495, 632) \\ 
&3&0.88 & 275 (240, 311) & 0.89 & 288 (251, 327) & 0.94 & 352 (306, 397) & 0.99 & 548 (478, 620) \\ 
\\[-1.8ex]  \hline\\[-1.8ex] 
\multirow{4}{*}{2}&  100&0.82 &  299 (262, 335) & 0.83 &  311 (272, 349) & 0.88 &  381 (332, 429) & 0.95 &  587 (516, 657)  \\ 
&5&0.83 &  291 (253, 326) & 0.83 &  300 (261, 337) & 0.88 &  370 (325, 413) & 0.94 &  567 (498, 636) \\ 
&4&0.82 &  286 (250, 322) & 0.83 &  295 (257, 332) & 0.88 &  363 (317, 407) & 0.94 &  559 (487, 629) \\ 
&3&0.83 &  274 (237, 310) & 0.84 &  283 (245, 322) & 0.88 &  347 (301, 394) & 0.93 &  541 (468, 611) \\ 
\\[-1.8ex]  \hline\\[-1.8ex] 
\multirow{4}{*}{3}&  100&0.75  & 192 (168, 215) & 0.75  & 197 (174, 219) & 0.79  & 225 (196, 250) & 0.88  & 291 (255, 325) \\ 
&5&0.76  & 187 (163, 209) & 0.76  & 190 (167, 212) & 0.81  & 217 (190, 243) & 0.88  & 281 (247, 314) \\
&4&0.77  & 183 (159, 206) & 0.78  & 187 (162, 210) & 0.79  & 212 (184, 239) & 0.88  & 277 (241, 309) \\
&3&  0.76  & 175 (150, 199) & 0.77  & 179 (154, 203) & 0.80  & 204 (176, 230) & 0.88  & 263 (226, 299) \\   
\\[-1.8ex]  \hline\\[-1.8ex]  
\multirow{4}{*}{4}&  100&0.71 &  191 (168, 213) & 0.71 &  197 (173, 220) & 0.73 &  223 (195, 249) & 0.82 &  289 (253, 323) \\ 
&5&0.72 &  186 (164, 208) & 0.72 &  189 (166, 211) & 0.75 &  216 (189, 242) & 0.82 &  278 (243, 312) \\ 
&4&0.73 &  183 (160, 205) & 0.71 &  186 (161, 209) & 0.76 &  212 (185, 238) & 0.82 &  272 (236, 307) \\ 
&3&0.71 &  174 (149, 198) & 0.72 &  177 (151, 202) & 0.76 &  202 (173, 230) & 0.82 &  261 (223, 297) \\ 
\\[-1.8ex] \hline \\[-1.8ex] 
\end{tabular} 
\end{sidewaystable}

\begin{sidewaystable} \centering 
\caption{Power and average sample size (first and third quartile in brackets) under Scenario 2 computed using the Bayesian `two priors' approach for different values of $\sigma_0$ and $\sigma_d$. The analysis prior mean $\theta_0$ is set to $0$.}
  \label{cont_bss_2_ni} 
  \small
\begin{tabular}{@{\extracolsep{-2pt}}ccccccccccc} 
\\[-1.8ex]\hline 
\hline \\[-2.7ex] 
&&&\multicolumn{8}{c}{$\sigma_d$}\\
&&\multicolumn{2}{c}{0}&\multicolumn{2}{c}{0.2}&\multicolumn{2}{c}{0.5}&\multicolumn{2}{c}{0.8}\\\cline{3-11}\\[-1.8ex]
Setting&$\sigma_0$& Power & $n$ & Power & $n$ &Power & $n$&Power & $n$ & \\
\hline\\[-1.8ex] 
\multirow{4}{*}{1}&    100& 0.88 & 741 (617, 856) & 0.87 & 769 (646, 880) & 0.93 & 942 (785, 1085) & 0.98 & 1452 (1221, 1671) \\
&3&0.88 & 767 (648, 882) & 0.88 & 785 (658, 907) & 0.92 & 963 (804, 1109) & 0.98 & 1493 (1249, 1726) \\
&2&0.87 & 782 (651, 905) & 0.88 & 810 (676, 933) & 0.93 & 990 (828, 1134) & 0.98 & 1516 (1269, 1750) \\
&1&0.86 & 899 (754, 1031) & 0.88 & 933 (780, 1077) & 0.92 & 1119 (935, 1291) & 0.98 & 1674 (1410, 1930) \\ 
\\[-1.8ex]  \hline\\[-1.8ex] 
\multirow{4}{*}{2}&  100&0.80 &  736 (615, 851) & 0.80 &  764 (638, 881) & 0.85 &  946 (791, 1089) & 0.90 &  1444 (1204, 1673) \\
&3&0.79 &  760 (630, 878) & 0.80 &  781 (655, 900) & 0.83 &  964 (806, 1118) & 0.90 &  1469 (1223, 1699) \\
&2&0.78 &  780 (647, 901) & 0.80 &  816 (682, 936) & 0.84 &  992 (836, 1138) & 0.91 &  1500 (1247, 1735) \\
&1&0.77 &  898 (752, 1027) & 0.78 &  928 (778, 1069) & 0.84 &  1126 (945, 1301) & 0.90 &  1665 (1391, 1923) \\
\\[-1.8ex]  \hline\\[-1.8ex] 
\multirow{4}{*}{3}&  100&0.74  & 473 (397, 544) & 0.73  & 484 (405, 556) & 0.80  & 551 (460, 637) & 0.85  & 713 (597, 822) \\
&3&0.74  & 496 (419, 568) & 0.73  & 506 (429, 580) & 0.79  & 577 (484, 662) & 0.87  & 746 (624, 863) \\
&2&0.73  & 520 (437, 598) & 0.72  & 531 (449, 609) & 0.79  & 599 (502, 690) & 0.86  & 768 (643, 883) \\
&1&  0.70  & 618 (519, 715) & 0.70  & 636 (535, 735) & 0.77  & 717 (598, 824) & 0.86  & 904 (761, 1039) \\   
\\[-1.8ex]  \hline\\[-1.8ex]  
\multirow{4}{*}{4}&  100&0.68 &  471 (391, 546) & 0.68 &  484 (401, 558) & 0.72 &  547 (456, 635) & 0.78 &  709 (587, 823) \\
&3& 0.67 &  493 (412, 567) & 0.68 &  504 (421, 580) & 0.72 &  571 (478, 664) & 0.79 &  740 (621, 853) \\
&2&0.68 &  518 (434, 594) & 0.68 &  528 (441, 610) & 0.70 &  595 (498, 684) & 0.79 &  768 (641, 886) \\
&1&0.65 &  616 (515, 710) & 0.65 &  633 (526, 731) & 0.69 &  707 (592, 815) & 0.77 &  890 (754, 1024) \\ 
\\[-1.8ex] \hline \\[-1.8ex] 
\end{tabular} 
\end{sidewaystable}

\begin{sidewaystable} \centering 
\caption{Power and average sample size (first and third quartile in brackets) under Scenario 2 computed using the Bayesian `two priors' approach for different values of $\sigma_0$ and $\sigma_d$. The analysis prior mean $\theta_0$ is set to $\widehat{\mu}^p_{\overbar{d}_k^1}-\widehat{\mu}^p_{\overbar{d}_k^2}$ (estimated from the simulated pilot study).}
  \label{cont_bss_2_i} 
  \small
\begin{tabular}{@{\extracolsep{-2pt}}ccccccccccc} 
\\[-1.8ex]\hline 
\hline \\[-2.7ex] 
&&&\multicolumn{8}{c}{$\sigma_d$}\\
&&\multicolumn{2}{c}{0}&\multicolumn{2}{c}{0.2}&\multicolumn{2}{c}{0.5}&\multicolumn{2}{c}{0.8}\\\cline{3-11}\\[-1.8ex]
Setting&$\sigma_0$& Power & $n$ & Power & $n$ &Power & $n$&Power & $n$ & \\
\hline\\[-1.8ex] 
\multirow{4}{*}{1}&    100& 0.88 & 739 (618, 853) & 0.87 & 765 (640, 883) & 0.93 & 942 (789, 1086) & 0.98 & 1459 (1224, 1673) \\ 
&5&0.88 & 723 (596, 837) & 0.88 & 746 (615, 870) & 0.92 & 919 (759, 1068) & 0.98 & 1419 (1187, 1637) \\ 
&4&0.87 & 716 (582, 843) & 0.88 & 737 (600, 861) & 0.93 & 904 (742, 1056) & 0.98 & 1402 (1159, 1635) \\ 
&3&0.86 & 685 (552, 814) & 0.88 & 716 (570, 846) & 0.92 & 876 (693, 1043) & 0.98 & 1353 (1087, 1593) \\  
\\[-1.8ex]  \hline\\[-1.8ex] 
\multirow{4}{*}{2}&  100&0.80 &  739 (623, 848) & 0.80 &  764 (635, 882) & 0.83 &  941 (782, 1088) & 0.91 &  1446 (1203, 1668) \\
&5&0.80 &  720 (590, 844) & 0.78 &  749 (616, 875) & 0.84 &  917 (752, 1070) & 0.90 &  1411 (1159, 1650) \\
&4&0.79 &  698 (563, 827) & 0.80 &  726 (584, 856) & 0.84 &  904 (726, 1069) & 0.90 &  1398 (1127, 1651) \\ 
&3&0.80 &  683 (539, 817) & 0.80 &  709 (563, 842) & 0.85 &  866 (678, 1032) & 0.90 &  1344 (1050, 1603) \\
\\[-1.8ex]  \hline\\[-1.8ex] 
\multirow{4}{*}{3}&  100&0.74  & 471 (396, 543) & 0.76  & 487 (405, 561) & 0.79  & 553 (461, 637) & 0.87  & 714 (602, 820) \\
&5&0.73  & 461 (378, 539) & 0.75  & 470 (384, 549) & 0.79  & 536 (439, 621) & 0.87  & 695 (572, 805) \\
&4&0.74  & 455 (368, 535) & 0.76  & 466 (378, 543) & 0.80  & 525 (428, 612) & 0.86  & 676 (548, 794) \\
&3&  0.75  & 431 (337, 519) & 0.76  & 449 (353, 534) & 0.79  & 507 (399, 602) & 0.87  & 656 (513, 785) \\
\\[-1.8ex]  \hline\\[-1.8ex]  
\multirow{4}{*}{4}&  100&0.70 &  474 (398, 546) & 0.71 &  483 (404, 554) & 0.71 &  545 (455, 632) & 0.80 &  714 (598, 823) \\
&5&0.69 &  456 (374, 533) & 0.71 &  469 (382, 550) & 0.72 &  531 (432, 623) & 0.78 &  685 (562, 801) \\
&4&0.70 &  450 (363, 530) & 0.72 &  463 (374, 544) & 0.74 &  522 (416, 617) & 0.79 &  678 (545, 800) \\
&3&0.71 &  434 (332, 522) & 0.72 &  447 (341, 540) & 0.74 &  509 (390, 617) & 0.78 &  659 (510, 787) \\
\\[-1.8ex] \hline \\[-1.8ex] 
\end{tabular} 
\end{sidewaystable}

\begin{table}[!htbp] \centering 
  \caption{Simulated power under the frequentist calculations in Scenarios 3 (left) and 4 (right) for different degrees of model misspecification.} 
  \label{power_res3_2} 
\begin{tabular}{@{\extracolsep{-1pt}} cccccccccccc} 
\\[-1.8ex]\hline \hline
&&&\multicolumn{5}{c}{Scenario 1} && \multicolumn{3}{c}{Scenario 2} \\\cline{4-8}\cline{10-12}\\[-1.8ex]
&&&\multicolumn{9}{c}{Response SD} \\
&& & 0 & 0.01 & 0.02 & 0.03 & 0.04  &&0 & 0.01 & 0.02 \\ 
\hline 
\multirow{5}{3em}{\% bias of $\delta$}&0 && 0.89 &0.88& 0.85& 0.84& 0.80 & & 0.83 &0.82& 0.79 \\ 
&5 && 0.87 &0.85 &0.84 &0.80 &0.77 & & 0.81 &0.80 &0.77\\ 
&10 && 0.85 &0.83& 0.81 &0.79 &0.74 && 0.78 &0.75& 0.74\\ 
&15 && 0.82 &0.80 &0.80 &0.76 &0.74 && 0.74 &0.71 &0.72\\ 
&20 && 0.78 &0.77 &0.75 &0.75 &0.72 && 0.70 &0.71 &0.69\\ 
&25 && 0.74 &0.74 &0.74 &0.72 &0.70 && 0.67 &0.68 &0.66 \\ 
\hline \\[-1.8ex] 
\end{tabular} 
\end{table} 

\begin{sidewaystable} \centering 
\caption{Power and average sample size (first and third quartile in brackets) under Scenario 3 computed using the Bayesian `two priors' approach for different values of $\sigma_0$ and $\sigma_d$. The analysis prior mean $\theta_0$ is set to $0$.}
  \label{bss_1_ni} 
  \small
\begin{tabular}{@{\extracolsep{-6pt}}ccccccccccccccc} 
\\[-1.8ex]\hline 
\hline \\[-2.7ex] 
&&&\multicolumn{12}{c}{$\sigma_d$}\\
&&\multicolumn{2}{c}{0}&\multicolumn{2}{c}{0.01}&\multicolumn{2}{c}{0.02}&\multicolumn{2}{c}{0.03}&\multicolumn{2}{c}{0.04}&\multicolumn{2}{c}{0.05}\\\cline{3-15}\\[-1.8ex]
Setting&$\sigma_0$& Power & $n$ & Power & $n$ &Power & $n$ &Power & $n$ &Power & $n$ &Power & $n$& \\
 \hline\\[-1.8ex] 
\multirow{4}{*}{1}&    5& 0.88 & 719 (652, 788) & 0.90 & 730 (661, 796) & 0.90 & 776 (700, 853) & 0.93 & 863 (786, 943) & 0.96 & 997 (909, 1094) & 0.98 & 1215 (1105, 1329)  \\ 
&0.5&0.89 & 724 (657, 795) & 0.90 & 740 (672, 809) & 0.90 & 784 (709, 860) & 0.93 & 864 (785, 943) & 0.96 & 995 (903, 1091) & 0.98 & 1219 (1107, 1339) \\ 
&0.3&0.89 & 735 (664, 806) & 0.90 & 747 (678, 817) & 0.90 & 788 (711, 865) & 0.93 & 871 (787, 955) & 0.96 & 1007 (911, 1106) & 0.98 & 1229 (1114, 1346) \\
&0.1&0.89 & 810 (732, 889) & 0.90 & 822 (748, 900) & 0.91 & 868 (787, 947) & 0.93 & 955 (869, 1044) & 0.96 & 1097 (992, 1204) & 0.98 & 1317 (1191, 1446) \\ 
\\[-1.8ex]  \hline\\[-1.8ex] 
\multirow{4}{*}{2}&  5&0.79 &  717 (647, 787) & 0.78 &  729 (658, 803) & 0.79 &  772 (696, 851) & 0.83 &  853 (770, 939) & 0.84 &  997 (898, 1099) & 0.88 &  1204 (1094, 1323) \\ 
&0.5&0.79 &  719 (654, 785) & 0.79 &  734 (662, 808) & 0.81 &  779 (700, 857) & 0.83 &  858 (780, 941) & 0.85 &  995 (900, 1096) & 0.88 &  1207 (1094, 1321) \\
&0.3&0.79 &  729 (658, 801) & 0.78 &  739 (666, 813) & 0.81 &  784 (708, 861) & 0.81 &  871 (789, 960) & 0.85 &  1002 (908, 1102) & 0.88 &  1222 (1110, 1341) \\ 
&0.1&0.78 &  803 (728, 881) & 0.78 &  816 (737, 898) & 0.79 &  859 (775, 941) & 0.82 &  951 (862, 1046) & 0.83 &  1087 (980, 1198) & 0.88 &  1316 (1189, 1441) \\
\\[-1.8ex]  \hline\\[-1.8ex] 
\multirow{4}{*}{3}&  5&0.75  & 462 (417, 505) & 0.75  & 466 (424, 512) & 0.76  & 483 (438, 528) & 0.78  & 517 (470, 566) & 0.82  & 565 (511, 619) & 0.86  & 634 (574, 695) \\  
&0.5&0.74  & 465 (420, 509) & 0.75  & 471 (427, 516) & 0.76  & 490 (443, 537) & 0.78  & 519 (469, 567) & 0.82  & 570 (516, 624) & 0.86  & 639 (577, 702) \\ 
&0.3&0.73  & 472 (426, 517) & 0.75  & 477 (433, 522) & 0.76  & 496 (450, 543) & 0.78  & 527 (478, 577) & 0.83  & 577 (525, 630) & 0.86  & 652 (592, 714) \\  
&0.1&  0.72  & 541 (491, 591) & 0.73  & 549 (500, 600) & 0.76  & 570 (520, 622) & 0.79  & 602 (546, 656) & 0.80  & 654 (592, 716) & 0.86  & 730 (662, 800) \\  
\\[-1.8ex]  \hline\\[-1.8ex]  
\multirow{4}{*}{4}  &5&0.69 &  458 (415, 503) & 0.68 &  466 (422, 510) & 0.70 &  480 (435, 527) & 0.71 &  515 (467, 567) & 0.74 &  561 (507, 616) & 0.75 &  631 (568, 696) \\
&0.5&0.69 &  464 (420, 509) & 0.69 &  468 (423, 513) & 0.70 &  485 (442, 530) & 0.71 &  520 (471, 569) & 0.73 &  565 (509, 623) & 0.78 &  637 (576, 701) \\ 
&0.3&0.69 &  469 (424, 515) & 0.67 &  475 (429, 523) & 0.70 &  492 (445, 540) & 0.71 &  526 (476, 577) & 0.73 &  572 (517, 628) & 0.76 &  646 (585, 705) \\
&0.1&0.66 &  540 (487, 592) & 0.68 &  546 (493, 599) & 0.67 &  564 (510, 621) & 0.69 &  597 (545, 653) & 0.74 &  655 (590, 722) & 0.75 &  726 (657, 797) \\ 
\\[-1.8ex] \hline \\[-1.8ex] 
\end{tabular} 
\end{sidewaystable}

\begin{sidewaystable} \centering 
\caption{Power and average sample size (first and third quartile in brackets) under Scenario 3 computed using the Bayesian `two priors' approach for different values of $\sigma_0$ and $\sigma_d$. The analysis prior mean $\theta_0$ is set to $\widehat{\mu}^p_{\overbar{d}_k^1}-\widehat{\mu}^p_{\overbar{d}_k^2}$ (estimated from the simulated pilot study).}
  \label{bss_1_i} 
  \small
\begin{tabular}{@{\extracolsep{-6pt}}ccccccccccccccc} 
\\[-1.8ex]\hline 
\hline \\[-2.7ex] 
&&&\multicolumn{12}{c}{$\sigma_d$}\\
&&\multicolumn{2}{c}{0}&\multicolumn{2}{c}{0.01}&\multicolumn{2}{c}{0.02}&\multicolumn{2}{c}{0.03}&\multicolumn{2}{c}{0.04}&\multicolumn{2}{c}{0.05}\\\cline{3-15}\\[-1.8ex]
Setting&$\sigma_0$& Power & $n$ & Power & $n$ &Power & $n$ &Power & $n$ &Power & $n$ &Power & $n$&  \\
 \hline\\[-1.8ex] 
\multirow{4}{*}{1}&    5& 0.89 & 719 (650, 790) & 0.90 & 734 (665, 804) & 0.91 & 777 (705, 852) & 0.93 & 862 (784, 945) & 0.96 & 996 (903, 1089) & 0.98 & 1218 (1105, 1334) \\ 
&0.5&0.89 & 711 (643, 784) & 0.89 & 722 (651, 799) & 0.91 & 770 (694, 849) & 0.93 & 850 (770, 935) & 0.96 & 984 (889, 1086) & 0.98 & 1199 (1079, 1321) \\ 
&0.3&0.89 & 697 (621, 780) & 0.90 & 709 (632, 791) & 0.92 & 752 (665, 839) & 0.93 & 824 (732, 920) & 0.95 & 962 (855, 1072) & 0.98 & 1175 (1047, 1310) \\
&0.2&0.89 & 661 (555, 767) & 0.90 & 678 (577, 786) & 0.91 & 721 (606, 841) & 0.93 & 786 (668, 913) & 0.96 & 908 (763, 1060) & 0.98 & 1114 (945, 1287) \\  
\\[-1.8ex]  \hline\\[-1.8ex] 
\multirow{4}{*}{2}&  5&0.78 &  718 (651, 787) & 0.80 &  731 (660, 803) & 0.79 &  773 (696, 849) & 0.81 &  853 (772, 934) & 0.84 &  985 (890, 1080) & 0.88 &  1208 (1092, 1327) \\
&0.5&0.79 &  709 (638, 781) & 0.79 &  717 (643, 791) & 0.80 &  763 (688, 840) & 0.82 &  847 (765, 935) & 0.84 &  980 (887, 1080) & 0.87 &  1192 (1071, 1315) \\
&0.3&0.79 &  690 (612, 774) & 0.79 &  706 (628, 789) & 0.81 &  747 (665, 838) & 0.82 &  822 (731, 918) & 0.84 &  955 (842, 1073) & 0.87 &  1164 (1036, 1302) \\ 
&0.2&0.79 &  657 (549, 766) & 0.79 &  671 (559, 787) & 0.80 &  714 (595, 836) & 0.82 &  780 (654, 916) & 0.86 &  907 (753, 1067) & 0.88 &  1111 (932, 1296) \\  
\\[-1.8ex]  \hline\\[-1.8ex] 
\multirow{4}{*}{3}&  5&0.74  & 461 (418, 504) & 0.76  & 468 (421, 515) & 0.75  & 485 (439, 531) & 0.79  & 518 (472, 568) & 0.82  & 565 (512, 617) & 0.85  & 637 (575, 700) \\   
&0.5&0.74  & 455 (408, 503) & 0.74  & 460 (413, 508) & 0.77  & 477 (429, 525) & 0.78  & 509 (458, 562) & 0.82  & 555 (500, 613) & 0.86  & 627 (565, 691) \\
&0.3&0.75  & 442 (391, 498) & 0.76  & 448 (394, 504) & 0.77  & 466 (410, 524) & 0.79  & 496 (439, 556) & 0.81  & 541 (476, 606) & 0.86  & 614 (544, 688) \\ 
&0.2&  0.76  & 417 (343, 495) & 0.75  & 424 (349, 502) & 0.77  & 438 (358, 519) & 0.79  & 466 (380, 555) & 0.82  & 511 (421, 607) & 0.86  & 573 (467, 683) \\ 
\\[-1.8ex]  \hline\\[-1.8ex]  
\multirow{4}{*}{4}&  5&0.67 &  459 (415, 501) & 0.68 &  466 (422, 511) & 0.70 &  482 (431, 530) & 0.72 &  513 (467, 563) & 0.74 &  557 (505, 612) & 0.76 &  636 (576, 694) \\ 
&0.5&0.66 &  453 (406, 502) & 0.69 &  459 (412, 507) & 0.70 &  476 (430, 524) & 0.71 &  506 (452, 557) & 0.74 &  555 (498, 612) & 0.76 &  623 (560, 688) \\
&0.3&0.69 &  440 (386, 496) & 0.69 &  446 (391, 503) & 0.70 &  461 (402, 523) & 0.71 &  491 (431, 555) & 0.74 &  536 (469, 604) & 0.77 &  607 (529, 686) \\ 
&0.2&0.70 &  417 (337, 504) & 0.71 &  419 (337, 506) & 0.72 &  435 (350, 523) & 0.73 &  463 (377, 554) & 0.75 &  507 (408, 607) & 0.78 &  576 (468, 689)\\ 
\\[-1.8ex] \hline \\[-1.8ex] 
\end{tabular} 
\end{sidewaystable}

\begin{sidewaystable} \centering 
\caption{Power and average sample size (first and third quartile in brackets) under Scenario 4 computed using the Bayesian `two priors' approach for different values of $\sigma_0$ and $\sigma_d$. The analysis prior mean $\theta_0$ is set to $0$.}
  \label{bss_2_ni} 
  \small
\begin{tabular}{@{\extracolsep{-2pt}}ccccccccccc} 
\\[-1.8ex]\hline 
\hline \\[-2.7ex] 
&&&\multicolumn{8}{c}{$\sigma_d$}\\
&&\multicolumn{2}{c}{0}&\multicolumn{2}{c}{0.01}&\multicolumn{2}{c}{0.02}&\multicolumn{2}{c}{0.03}\\\cline{3-11}\\[-1.8ex]
Setting&$\sigma_0$& Power & $n$ & Power & $n$ &Power & $n$&Power & $n$&  \\
\hline\\[-1.8ex] 
\multirow{4}{*}{1}&    5& 0.87 & 1843 (1532, 2191) & 0.89 & 1975 (1646, 2345) & 0.93 & 2443 (2028, 2903) & 0.98 & 3637 (3045, 4306) \\ 
&0.5&0.87 & 1849 (1542, 2197) & 0.89 & 1981 (1659, 2346) & 0.93 & 2438 (2027, 2906) & 0.98 & 3658 (3035, 4343) \\ 
&0.3&0.87 & 1856 (1541, 2206) & 0.89 & 1978 (1644, 2358) & 0.93 & 2447 (2036, 2904) & 0.98 & 3636 (3025, 4317) \\
&0.1&0.86 & 1920 (1607, 2273) & 0.90 & 2060 (1709, 2446) & 0.93 & 2516 (2093, 2987) & 0.98 & 3742 (3100, 4442) \\
\\[-1.8ex]  \hline\\[-1.8ex] 
\multirow{4}{*}{2}&  5&0.83 &  1841 (1528, 2189) & 0.84 &  1971 (1645, 2338) & 0.89 &  2448 (2035, 2906) & 0.95 &  3622 (3028, 4300) \\  
&0.5&0.82 &  1846 (1540, 2190) & 0.84 &  1973 (1650, 2342) & 0.88 &  2444 (2046, 2898) & 0.95 &  3644 (3019, 4326) \\
&0.3&0.82 &  1860 (1546, 2208) & 0.84 &  1983 (1660, 2352) & 0.88 &  2447 (2041, 2898) & 0.94 &  3632 (3028, 4303) \\
&0.1&0.82 &  1912 (1598, 2270) & 0.85 &  2052 (1700, 2436) & 0.88 &  2512 (2095, 2977) & 0.95 &  3720 (3074, 4408) \\  
\\[-1.8ex]  \hline\\[-1.8ex] 
\multirow{4}{*}{3}&  5&0.76  & 1182 (982, 1396) & 0.74  & 1227 (1022, 1454) & 0.79  & 1402 (1170, 1666) & 0.86  & 1792 (1493, 2129) \\ 
&0.5&0.71  & 1167 (974, 1388) & 0.75  & 1236 (1029, 1463) & 0.79  & 1415 (1178, 1677) & 0.86  & 1784 (1486, 2112) \\
&0.3&0.74  & 1194 (996, 1411) & 0.74  & 1240 (1026, 1472) & 0.80  & 1426 (1193, 1677) & 0.86  & 1790 (1485, 2120) \\ 
&0.1&  0.71  & 1250 (1045, 1476) & 0.76  & 1303 (1080, 1548) & 0.80  & 1479 (1236, 1743) & 0.86  & 1869 (1560, 2206) \\   
\\[-1.8ex]  \hline\\[-1.8ex] 
\multirow{4}{*}{4}&  5&0.70 &  1190 (982, 1414) & 0.72 &  1242 (1026, 1468) & 0.77 &  1422 (1182, 1680) & 0.81 &  1796 (1488, 2130) \\
&0.5&0.70 &  1183 (979, 1406) & 0.72 &  1237 (1027, 1460) & 0.75 &  1410 (1170, 1666) & 0.82 &  1795 (1484, 2139) \\
&0.3&0.70 &  1192 (987, 1416) & 0.72 &  1249 (1037, 1476) & 0.76 &  1410 (1174, 1673) & 0.82 &  1790 (1482, 2124) \\ 
&0.1&0.70 &  1248 (1040, 1476) & 0.70 &  1297 (1080, 1531) & 0.74 &  1498 (1248, 1770) & 0.83 &  1854 (1529, 2210) \\ 
\\[-1.8ex] \hline \\[-1.8ex] 
\end{tabular} 
\end{sidewaystable}

\begin{sidewaystable} \centering 
\caption{Power and average sample size (first and third quartile in brackets) under Scenario 4 computed using the Bayesian `two priors' approach for different values of $\sigma_0$ and $\sigma_d$. The analysis prior mean $\theta_0$ is set to $\widehat{\mu}^p_{\overbar{d}_k^1}-\widehat{\mu}^p_{\overbar{d}_k^2}$ (estimated from the simulated pilot study).}
  \label{bss_2_i} 
  \small
\begin{tabular}{@{\extracolsep{-2pt}}ccccccccccc} 
\\[-1.8ex]\hline 
\hline \\[-2.7ex] 
&&&\multicolumn{8}{c}{$\sigma_d$}\\
&&\multicolumn{2}{c}{0}&\multicolumn{2}{c}{0.01}&\multicolumn{2}{c}{0.02}&\multicolumn{2}{c}{0.03}\\\cline{3-11}\\[-1.8ex]
Setting&$\sigma_0$& Power & $n$ & Power & $n$ &Power & $n$&Power & $n$ & \\
\hline\\[-1.8ex] 
\multirow{4}{*}{1}&    5& 0.87 & 1843 (1532, 2191) & 0.89 & 1975 (1646, 2345) & 0.93 & 2443 (2028, 2902) & 0.98 & 3637 (3045, 4306) \\ 
&0.5&0.87 & 1840 (1533, 2187) & 0.89 & 1971 (1651, 2334) & 0.93 & 2425 (2016, 2888) & 0.98 & 3641 (3023, 4329) \\
&0.3&0.88 & 1830 (1520, 2183) & 0.89 & 1948 (1620, 2325) & 0.93 & 2413 (2013, 2866) & 0.98 & 3590 (2987, 4262) \\ 
&0.2&0.87 & 1807 (1502, 2145) & 0.89 & 1940 (1600, 2315) & 0.94 & 2380 (1952, 2829) & 0.98 & 3572 (2946, 4260) \\
\\[-1.8ex]  \hline\\[-1.8ex] 
\multirow{4}{*}{2}&  5&0.83 &  1840 (1528, 2189) & 0.84 &  1971 (1644, 2338) & 0.89 &  2448 (2035, 2906) & 0.95 &  3621 (3028, 4299) \\  
&0.5&0.83 &  1836 (1536, 2182) & 0.84 &  1963 (1644, 2333) & 0.88 &  2431 (2035, 2886) & 0.94 &  3627 (3011, 4302) \\ 
&0.3&0.83 &  1833 (1525, 2183) & 0.84 &  1953 (1637, 2322) & 0.88 &  2413 (2016, 2863) & 0.94 &  3584 (2988, 4249) \\ 
&0.2&0.83 &  1798 (1488, 2141) & 0.85 &  1933 (1586, 2308) & 0.89 &  2375 (1955, 2823) & 0.95 &  3550 (2910, 4224) \\  
\\[-1.8ex]  \hline\\[-1.8ex] 
\multirow{4}{*}{3}&  5&0.74  & 1181 (980, 1391) & 0.74  & 1241 (1024, 1472) & 0.79  & 1409 (1178, 1657) & 0.86  & 1784 (1483, 2107) \\  
&0.5&0.73  & 1177 (981, 1394) & 0.75  & 1237 (1024, 1468) & 0.79  & 1413 (1169, 1676) & 0.87  & 1781 (1475, 2119) \\ 
&0.3&0.74  & 1179 (969, 1413) & 0.74  & 1219 (1011, 1453) & 0.79  & 1383 (1143, 1653) & 0.86  & 1762 (1469, 2090)  \\ 
&0.2&  0.73  & 1149 (937, 1362) & 0.75  & 1197 (962, 1432) & 0.80  & 1366 (1100, 1621) & 0.86  & 1731 (1411, 2061) \\    
\\[-1.8ex]  \hline\\[-1.8ex]  
\multirow{4}{*}{4}&  5&0.71 &  1183 (980, 1412) & 0.71 &  1234 (1023, 1467) & 0.75 &  1414 (1180, 1666) & 0.82 &  1780 (1478, 2103) \\ 
&0.5&0.71 &  1189 (986, 1407) & 0.72 &  1232 (1016, 1462) & 0.77 &  1404 (1163, 1671) & 0.81 &  1771 (1466, 2100) \\ 
&0.3&0.70 &  1168 (970, 1384) & 0.72 &  1212 (1021, 1435) & 0.76 &  1390 (1151, 1652) & 0.82 &  1761 (1474, 2091) \\ 
&0.2&0.71 &  1144 (918, 1369) & 0.71 &  1197 (969, 1422) & 0.76 &  1367 (1093, 1629) & 0.81 &  1731 (1416, 2059) \\ 
\\[-1.8ex] \hline \\[-1.8ex] 
\end{tabular} 
\end{sidewaystable}

\begin{table}[!htbp] \centering 
  \caption{Simulated type I error under the proposed methodology for different choices of prior parameters in the binary final outcome setting.} 
  \label{alpha_bin} 
\begin{tabular}{@{\extracolsep{6pt}} cccccccccc} 
\\[-1.8ex]\hline 
\hline 
&&&&\multicolumn{6}{c}{$\sigma_d$} \\
&&& & 0 & 0.01 & 0.02 & 0.03 & 0.04 & 0.05 \\ 
\hline \\[-1.8ex] 
\multirow{4}{3.2em}{$\theta_0=0$}&\multirow{4}{1em}{$\sigma_0$}&5&& $0.052$ & $0.050$ & $0.055$ & $0.057$ & $0.055$ & $0.050$ \\ 
&&0.5 && $0.060$ & $0.048$ & $0.045$ & $0.053$ & $0.052$ & $0.058$ \\ 
&&0.3& & $0.045$ & $0.044$ & $0.048$ & $0.052$ & $0.048$ & $0.039$  \\ 
&&0.1 && $0.035$ & $0.034$ & $0.040$ & $0.040$ & $0.038$ & $0.042$ \\ 
\\[-1.8ex] 
\hline\\[-1.8ex] 
\multirow{4}{3.2em}{$\theta_0=\widehat{\theta}^p$}&\multirow{4}{1em}{$\sigma_0$}&5 && $0.053$ & $0.054$ & $0.047$ & $0.046$ & $0.051$ & $0.049$ \\ 
&&0.5 && $0.061$ & $0.051$ & $0.042$ & $0.059$ & $0.049$ & $0.053$ \\ 
&&0.3 && $0.054$ & $0.051$ & $0.052$ & $0.054$ & $0.049$ & $0.042$ \\ 
&&0.2 && $0.049$ & $0.056$ & $0.055$ & $0.053$ & $0.054$ & $0.060$ \\ 
\hline \\[-1.8ex] 
\end{tabular} 

\end{table}

\label{lastpage}